\documentclass[journal=mamobx, articletitle=true]{achemso}
\setkeys{acs}{articletitle = true}

\usepackage{graphicx}
\usepackage{amsmath}
\usepackage{caption}
\usepackage{subcaption}
\usepackage{amsthm,amssymb,enumerate}%, esint}
\usepackage{mathtools}
\usepackage{color}
\usepackage{bm}
\usepackage{paracol}
\usepackage{changes}

\title{Viscosity of flexible and semiflexible ring melts - molecular origins and flow-induced segregation}

\author{Ranajay Datta}
\affiliation{Institute of Physics, Johannes Gutenberg University, Staudingerweg 9, 55128 Mainz, Germany}

\author{Fabian Berressem}
\affiliation{Institute of Physics, Johannes Gutenberg University, Staudingerweg 9, 55128 Mainz, Germany}

\author{Friederike Schmid}
\affiliation{Institute of Physics, Johannes Gutenberg University, Staudingerweg 9, 55128 Mainz, Germany}

\author{Arash Nikoubashman}
\affiliation{Institute of Physics, Johannes Gutenberg University, Staudingerweg 9, 55128 Mainz, Germany}

\author{Peter Virnau}
\email{virnau@uni-mainz.de}
\affiliation{Institute of Physics, Johannes Gutenberg University, Staudingerweg 9, 55128 Mainz, Germany}

\date{\today}

\begin{document}
\begin{abstract}
We investigate with numerical simulations the molecular origin of viscosity in melts of flexible and semiflexible oligomer rings in comparison to corresponding systems with linear chains. The strong increase of viscosity with ring stiffness is linked to the formation of entangled clusters, which dissolve under shear. This shear-induced breakup and alignment of rings in the flow direction lead to pronounced shear-thinning and non-Newtonian behavior. In melts of linear chains, the viscosity can be associated with the (average) number of entanglements between chains, which also dissolve under shear. While blends of flexible and semiflexible rings are mixed at rest, the two species separate under flow. This phenomenon has potential applications in microfluidic devices to segregate ring polymers of similar mass and chemical composition by their bending rigidity.

\end{abstract}

\maketitle
% Keywords
%\keyword{} 

\section{Introduction}
\label{sec_1}

Ring polymers occur abundantly in biological and synthetic polymers \cite{Trabi_2002, Sanchez_2010, Cristian_2012, Vinogradov_2019, Wu_2017, Kaitz_2013}. Over the past decades, numerous naturally occurring circular protein species have been identified and studied, and biochemical techniques to artificially synthesize various species of circular proteins have been developed \cite{Trabi_2002}. It has been established that the DNA of various living organisms, e.g., the polyoma virus or cytoplasmic DNA of animals, exist as circular rings \cite{vologodskii_2015}. Many of these naturally occurring or artificially synthesized ring polymers are rather useful for a number of biochemical processes: For example, macrocyclic peptides exhibit a number of advantageous pharmacological properties, which make them uniquely effective in drug modalities \cite{Vinogradov_2019}. In a good solvent, the equilibrium conformation of a ring polymer strongly depends on the stiffness of its segments, with coil-like configurations for the fully flexible case and circular shapes at large bending stiffness \cite{Karen_2007, Camacho_1991}. Understanding and predicting the conformational changes of ring polymers exposed to non-equilibrium processes like shear is also important for various biochemical processes like the synthesis and processing of plasmid DNA \cite{Levy_2000}, which in turn have applications in the design of novel vaccines and gene therapy, and the motion of DNA through nanochannels \cite{Sheng_2012}. A comprehensive understanding of the interdependence of molecular conformations of flexible and semiflexible ring polymers exposed to shear and their rheological properties like viscosity is thus of paramount importance. 
%\textcolor{red}{ It is important to note that polymer architecture and topology impact rheological properties. Numerous studies \cite{Singh_2013, Padding_2010, Jabbarzadeh_2003} have been conducted that study the rheological properties of branched polymers like H, comb and star polymers. They have studied in detail how topology and architecture typical to the different types of branched polymers contribute to them having different rheological properties than ring polymers.}

Over the years, many rheological properties of
ring polymer melts have been studied in simulations and experiments \cite{Halverson_2011, Pasquino_2013, Tsamopolous_2019, Murashima_2021, Parisi_2021}. It has been observed \cite{Pasquino_2013} that in the regime of very low entanglements, viscosities pertaining to linear and ring polymers exhibit a similar scaling behavior with respect to entanglements, resulting in the ratio of viscosities of ring and linear polymeric systems being almost constant. However, in the strongly entangled regime, the scaling behavior of the two systems is evidently different \cite{Pasquino_2013}. 
Moreover, the scaling relation of the zero-shear viscosity as a function of molecular weight is much weaker for ring polymer melts compared to linear melts \cite{Halverson_2011},
and melts of flexible entangled rings \cite{Parisi_2021} also exhibit weaker shear thinning.
%as compared to linear entangled melts. 
%MD simulations of melts of large rings \cite{Halverson_2011} exhibit differences in the diffusive behavior of rings in melts compared to their linear counterpart. 
%Moreover, the scaling relation of the zero shear viscosity as a function of N is much weaker for large ring polymer melts compared to linear melts \cite{Halverson_2011}. 
%Studies examining shear rheology of flexible, unentangled or slightly entangled ring polymer melts \cite{Tsamopolous_2019} reveal, quite unexpectedly, that ring-ring threading events survive even under high shear.

Some studies have also investigated rheological properties of shorter, semiflexible linear \cite{Datta_Virnau_2021,Xu_2017, nikoubashman:mm:2017, Kong2019, Winkler_2006, Winkler_2010, Huang_Gompper_Winkler_2012, Huang_Gommper_Winkler_2012_2} and ring polymers \cite{Hsiao_2016, Young_2019, Cifre_2005, chen_2013} 
%and their conformational response to shear 
in solution revealing transitions between stretching, coiling, and tumbling motion depending on the specific flow conditions.
For example, Chen \textit{et al.} \cite{chen_2013} carried out a comprehensive study on individual semiflexible ring polymers under shear, covering a large range of Weissenberg numbers, using molecular dynamics (MD) simulations coupled to the multi-particle collision dynamics (MPCD) scheme \cite{Winkler_2005} to include hydrodynamic interactions. They conducted simulations over a range of chain stiffnesses and reported how the latter affects conformations, dynamical modes and rheological properties of rings under shear. 

Recently, numerical investigations of more concentrated solutions of shorter, semiflexible rings have also gained attention\cite{Bernabei_2013, Slimani_2014, Poier_2015, Poier_2016, avendano:pnas:2016, Weiss_Nikoubashman_2019, Likos_Liebetreu_2020, roy:arxiv:2023}, which exhibit cluster phases, wherein semiflexible rings can interpenetrate each other and form stacks. 
Liebetreu \textit{et al.} \cite{Likos_Liebetreu_2020} conducted MPCD \cite{Winkler_2005} simulations to model solutions of semiflexible rings (at low to medium concentrations) under shear. They quantified the degree of clustering in the solutions and also demonstrated quantitatively that an increase in the shear rate is associated with an alignment of the chains in the flow direction accompanied by a progressive breakdown of clusters. We complement those previous studies by investigating the influence of chain stiffness on the shear viscosity of highly concentrated melts. We find that the melts become more viscous with increasing bending stiffness, and associate this behavior to the formation of stacked ring clusters. In contrast, the rheological properties of melts of linear chains at zero shear are dominated by a different molecular mechanism, namely emergent entanglements in the semiflexible regime \cite{Datta_Virnau_2021}, which we quantify with the Z1 algorithm \cite{Kroger_2005, Karayiannis_2009, Hoy_2009}.

In the final section of our work, we study a binary mixture of flexible and stiff rings. We determine zero-shear viscosities and explain the composition dependence of mixture viscosities by estimating the degree of clustering as a function of composition. While stiff and flexible rings form a homogeneous mixture in equilibrium, the two species can become segregated in flow due to differences in rheological properties of the individual components. Differences in rheological properties emerging from topology have also been observed for branched polymers like H-, comb or star polymers \cite{Singh_2013, Padding_2010, Jabbarzadeh_2003} and could in principle lead to similar effects. Indeed, flow-induced segregation was observed previously also for mixtures of flexible ring and chain polymers \cite{Weiss_Nikoubashman_2019} and mixtures of flexible chains and stars \cite{Srivastava_Nikoubashman_2018}. 
Flow induced segregation in mixtures of particles having different mechanical properties like elasticity have also been reported in the literature \cite{Muller_2014, Guu_2013, Kumar_2012} and are of particular relevance in the context of blood flow \cite{Muller_2014, Kumar_2012}.
Here, we demonstrate that ring polymers of similar mass and chemical composition can in principle be separated in microfluidic devices, even if they do not phase separate in equilibrium.

\section{Microscopic Model and Simulation Techniques}
\label{sec_2}

In our microscopic model, oligomers are represented by standard bead-spring chains, following the formulations of Kremer and Grest \cite{Kremer1990}. Each pair of beads interacts via a repulsive Weeks-Chandler-Andersen (WCA) potential \cite{Weeks_1971}:

\begin{equation}
\begin{split}
V_{\rm WCA}(r)&=4\varepsilon\left [\left( \frac{\sigma}{r}\right) ^{12}-\left(\frac{\sigma}{r}\right)^{6} + \frac{1}{4}\right],\hspace{1cm}r\leq2^{1/6}\sigma \\
&=0,\hspace{5.3cm}r>2^{1/6}\sigma
\label{WCA}
\end{split}
    \end{equation}
with bead diameter $\sigma$ and interaction strength $\varepsilon$, which will be taken as the units of length and energy, respectively (units will be omitted in the following sections for brevity and all quantities will be expressed in simulation units). Adjacent beads, in addition to the WCA potential are connected with finitely extensible nonlinear elastic (FENE) bonds \cite{FENE}:

\begin{equation}
    V_{\rm FENE}=-\frac{1}{2}KR^{2}\ln\left[1-\left(\frac{r}{R}\right)^{2}\right]
\end{equation}
with spring constant $K=30$ and maximum bond extension $R=1.5$. 

Bending rigidity for the semiflexible chains is implemented by introducing a bending potential:

\begin{equation}
    V_\theta=\kappa(1+\cos{\theta})
    \label{bending}
\end{equation}
where $\theta$ is the angle between three consecutive beads and $\kappa$ is the stiffness coefficient. A cosine type bending potential like Eq.~\eqref{bending} owes its origin to the well-known Kratky-Porod model \cite{Kratky_49, DoiEdwards, RubinsteinColby} and is quite commonly used to model semiflexibility in polymers \cite{Auhl2003jcp, milchev:jcp:2018, nikoubashman:jcp:2021}. For sufficiently large interaction strengths $\kappa \gg 1$, the persistence length can be approximated by $\ell_{\rm p}/\ell_{\rm b} \approx \kappa$, with bond length $\ell_{\rm b} \approx \sigma$ in this model, as expected from the equipartition theorem.\cite{milchev:jcp:2018, nikoubashman:jcp:2021}

We perform non-equilibrium MD simulations of sheared oligomer melts at a monomer number density $\rho=0.8$ using the LAMMPS package \cite{Plimpton1995}. System sizes are $15^3$ unless stated otherwise explicitly. The flow direction (f) of shear is parallel to the $x$-axis, the gradient direction (g) is along the $y$-axis, and the vorticity direction (v) is along the $z$-axis of the simulation box. We introduce shear by superimposing a velocity gradient on the thermal velocities of the particles by virtue of the SLLOD equations \cite{Evans1984, Ladd1984, Tuckerman1997, Evans2008}. The latter essentially modify the equations of motion by adding a height-dependent velocity component and represent one of the standard microscopic approaches to model shear. The SLLOD equations were coupled to a Nos{\'e}-Hoover thermostat \cite{Evans1985, Tuckerman1997} to maintain a constant temperature $T=1$ throughout our simulations. The equations of motion were integrated using the Velocity Verlet algorithm. The simulation box that LAMMPS implements is non-orthogonal with periodic boundary conditions and undergoes continuous deformation in accordance with the external shear \cite{Evans1979, Hansen1994}. It has been established \cite{Evans2008, Todd2017} that this approach is equivalent to the commonly used Lees-Edwards boundary conditions. We calculated the shear viscosity $\eta(\dot{\gamma})$ using the relation

\begin{equation}
\eta=\frac{\sigma_{xy}}{\dot{\gamma}},
\end{equation} 
with $\dot{\gamma}$ being the applied shear rate and $\sigma_{xy}$ a non-diagonal component of the stress tensor, as determined by the Irving-Kirkwood formula \cite{Irving1950,allen-tildesley-87}:
\begin{equation}
  \sigma_{xy}=-\frac{1}{V}\left[\sum_{i}^{N}\left(m_{i}v_{i,x}v_{i,y}\right)+ \sum_{i}^{N}\sum_{j>i}^{N}\left(r_{ij,x}f_{ij,y}  \right)    \right].
\label{vis}
\end{equation}
Here, $V$ is the volume of the system, $m_{i}$, $\bf{v}_{i}$ are mass and peculiar velocity of the $i^\text{th}$ particle respectively, and $\bf{r}_{ij}$ is the distance vector and $\bf{f}_{ij}$ the force vector between the $i^\text{th}$ and the $j^\text{th}$ particle. As reference for small shear rates $\dot{\gamma}\rightarrow{0}$, we have also calculated the zero-shear viscosity {\it via} the Green-Kubo (GK) relation:

\begin{equation}
\eta_{\rm GK}=\frac{V}{k_{\rm B}T}\int_{0}^{\infty}\left\langle\sigma_{xy}(t)\sigma_{xy}(0)\right\rangle dt
\label{GK}
\end{equation}
with Boltzmann constant $k_{\rm B}$.
%
%\textcolor{red}{%
%Note that forces arising from the thermostat and its coupling to the SLLOD conditions are not explicitly considered in Eq.~\eqref{vis} and may potentially result in a small systematic error. For detailed discussion of this effect in the context of dissipative thermostats the reader is referred to Ref.~\citenum{Jung_2016}.
%}%
Note that Eq.~\eqref{vis} does not take into account forces that arise due to the thermostat and its connection to the SLLOD conditions, which might lead to a minor systematic error \cite{Jung_2016}. 

In the final section of our work, we simulate pressure-driven flow in a slit-like channel containing a binary mixture of polymers with stiffness $\kappa =0$ and $10$. Channel dimensions are $15\times30\times15$, and each channel wall consists of atoms arranged in a face-centered-cubic configuration with particle number density of 4.0 \cite{Duan_2015}. Wall particles are tethered to their respective positions and do not exert any force on each other. Polymer beads interact with the wall particles through a WCA potential with slightly smaller diameter 
%\begin{equation}
%\begin{split}
%V_{\rm wp}(r)&=4\epsilon_{wp}\left [\left( \frac{\sigma_{wp}}{r}\right) %^{12}-\left(\frac{\sigma_{wp}}{r}\right)^{6} + \frac{1}{4}\right],\hspace{1cm}r<2^{1/6} \\
%&=0,\hspace{6.4cm}r>2^{1/6}
%\end{split}
%\end{equation}
$\sigma_\text{wp}=0.85$  \cite{Duan_2015} and stronger repulsion $\varepsilon_\text{wp}=4.0$. A temperature of $T=1$ is maintained across the channel by using the dissipative particle dynamics (DPD) thermostat \cite{Soddemann_Duenweg_2003, Pastorino_Binder_2007, Binder_DPD_2011}, as implemented in LAMMPS. The DPD friction coefficient, $\lambda_\text{DPD}$, is chosen to be 4.5 and the cutoff distance $r_\text{c,DPD}$ for dissipative and random forces pertaining to the DPD thermostat is chosen to be $2\times2^{1/6}$ \cite{Pastorino_confproc_2015, Pastorino_2014}, while the cutoff for conservative WCA interactions between the polymer beads remains $2^{1/6}$.

\newpage

\section{Viscosity and Shear-thinning in Melts of Oligomer Rings as a Function of Stiffness}
\label{sec_3}

In the following section, we analyze how stretching, alignment and cluster formation in ring polymer melts govern macroscopic rheological properties, and compare results with linear oligomers \cite{Datta_Virnau_2021} as a function of chain stiffness.

 \begin{figure}[ht!]
      \includegraphics[width=3.1in]{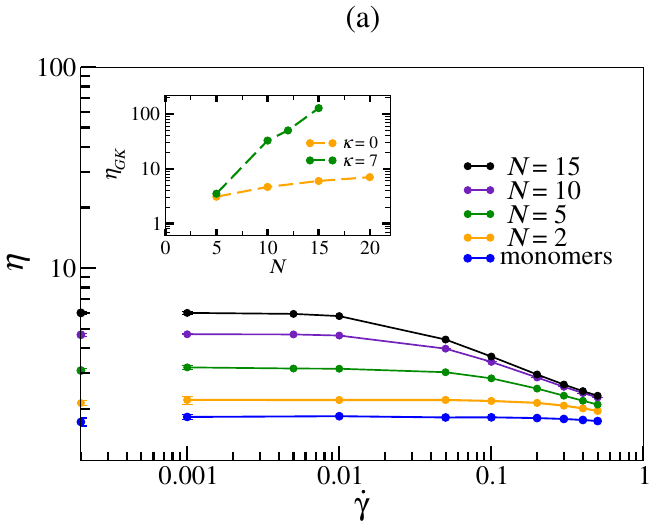}
      \hspace{0.3in}
      \includegraphics[width=3.0in]{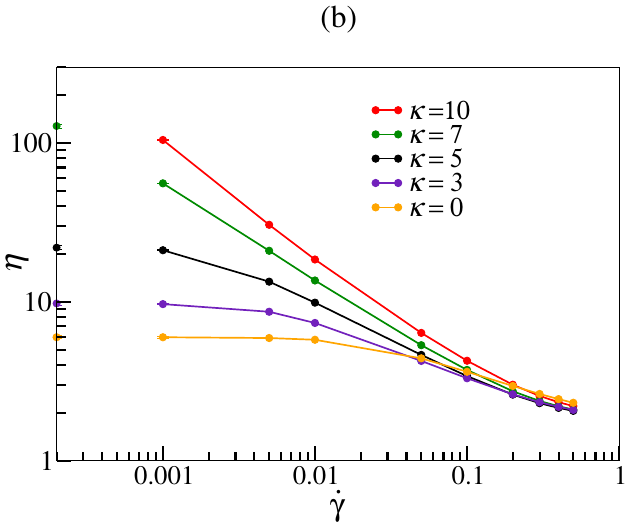}
    \caption{ \textbf{(a)} Shear viscosity $\eta$ as function of shear rate $\dot{\gamma}$ for melts of monomers, flexible dimers and ring polymers ($\kappa=0$) with $N= 5, 10$ and 15 beads per chain. Corresponding shear viscosities according to the Green-Kubo relation $\eta_\text{GK}$ are shown on the $y$-axis. $\eta_{GK}$ as a function of $N$ for ring polymers with $\kappa = 0$ and 7 is shown as an inset. Density $\rho=0.8$ and box dimensions are $10\times10\times10$ for $N= 1$ and 2,  $15\times15\times15$ for $N=5, 10, 15$ and $20\times20\times20$ for $N=20$.
\textbf{(b)} Shear viscosity $\eta$ as function of $\dot{\gamma}$ for melts of ring polymers with varying stiffness $\kappa$ at fixed $N= 15$.} 
\label{fig1}
\end{figure}

Figure~\ref{fig1}a displays shear viscosities $\eta$ for different degrees of polymerization $N$ as a function of shear rate $\dot{\gamma}$. The curves have a qualitatively similar progression as corresponding data for flexible linear polymers \cite{Datta_Virnau_2021}: $\eta$ is constant at small $\dot{\gamma}$ and increases with increasing $N$. The shear viscosity then drops with increasing $\dot{\gamma}$, where the onset of shear thinning shifts to smaller $\dot{\gamma}$ for larger $N$, as expected, since longer rings relax more slowly than short ones \cite{Weiss_Nikoubashman_2019}. We also plot the zero-shear viscosities as determined by the Green-Kubo relation, $\eta_{\rm GK}$, for flexible ring polymers of different sizes on the $y$-axis of Fig.~\ref{fig1}a. Note that $\eta_{\rm GK}$ values are consistent with $\eta(\dot{\gamma}\rightarrow{0})$. While for flexible rings, $\eta_{\rm GK}$ increases linearly with $N$, for $\kappa=7$ we observe an exponential growth as displayed in the inset of Fig.~\ref{fig1}a. Figure~\ref{fig1}b shows $\eta$ as a function of  $\dot{\gamma}$ for ring polymers of size $N=15$ and stiffness ranging from $\kappa=0$ (flexible rings) to $\kappa=10$ (stiffer rings). Again, we observe that the zero-shear viscosity $\eta_{\rm GK}$ and the viscosity $\eta(\kappa)$ at low shear rates exhibit a steep increase with increasing $\kappa$. This behavior is in contrast to that of linear polymers, where $\eta_{\rm GK}$ and $\eta$ at low shear rates display a non-monotonic dependence on $\kappa$,\cite{Datta_Virnau_2021} which we will further illustrate and explain in the next paragraph.

\begin{figure}[ht!]
%\centering

      \includegraphics[width=3.2in]{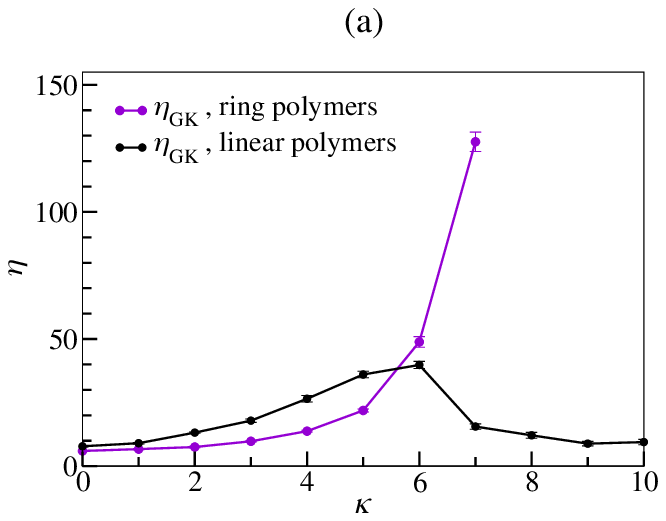}
      \includegraphics[width=3.2in]{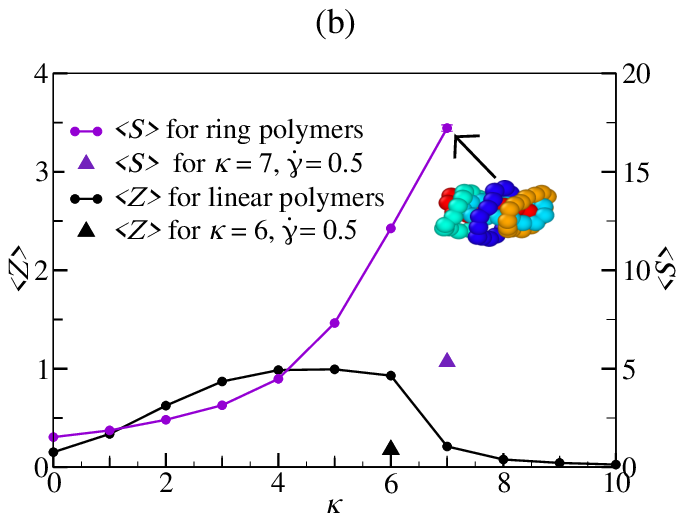}
 
\caption{
\textbf{(a)} Zero-shear viscosity from the Green-Kubo relation $\eta_\text{GK}$ as a function of $\kappa$ for linear and ring polymers ($N=15$ monomers).
\textbf{(b)} For the same systems (at equilibrium): Average number of entanglements per chain $\langle Z \rangle$ (left $y$-axis) as a function $\kappa$ for linear polymers and average number of ring polymers $\langle S \rangle$ per cluster (right $y$-axis) as a function of $\kappa$. The two triangles represent $\langle Z \rangle$ (at $\kappa=6$) and $\langle S \rangle$ (at $\kappa=7$) for high shear rates ($\dot{\gamma}=0.5)$.}
\label{fig2}
\end{figure}

Figure \ref{fig2}a shows $\eta_{\rm GK}$ as functions of $\kappa$ for semiflexible linear and ring polymers of length $N= 15$. For linear chains, $\eta_{\rm GK}$ initially increases with increasing $\kappa$, but then decreases for $\kappa > 6$. In our previous work \cite{Datta_Virnau_2021}, we have rationalized this non-monotonic behavior qualitatively as follows: The initial increase in $\eta_{\rm GK}$ was attributed to an increase in entanglements with increasing $\kappa$  \cite{Datta_Virnau_2021, Faller_CPC_2001, Faller1999, Hoy_2020}; however, near $\kappa =6$, the melt undergoes an isotropic-nematic transition \cite{Datta_Virnau_2021, Xu_2017}, resulting in fewer entanglements between the aligned chains accompanied by a sharp drop in $\eta_{\rm GK}$. In order to quantify this behavior, we now compute the average number of entanglements per chain, $\left\langle Z \right\rangle$ as a function of $\kappa$, by employing the Z1 code \cite{Kroger_2005, Karayiannis_2009, Hoy_2009, Shanbhag_2007}. This algorithm 
%(based on the Primitive Path Analysis (PPA) method) 
carries out a geometric minimization of the contour length of polymer chains to estimate the so-called primitive paths, which finally leads to an enumeration of entanglements per chain. 
%The primitive path of a chain in the melt is constructed by first representing the backbone of the chain as an infinitely thin series of multiple disconnected paths. 
%The length of the multiple disconnected path is then reduced at each step, by decreasing the number of nodes, 
Essentially, the Z1 algorithm successively removes all monomers which do not contribute to interchain entanglements. This is achieved by keeping a monomer between two consecutive segments if and only if a neighbouring polymer intersects the plane formed by the two consecutive segments. In this case, the monomer in the middle is also moved towards the entanglement. The process is repeated until the final primitive path conformation has been achieved. The number of kinks in the primitive path is used to estimate the number of entanglements encountered by that chain. 
%Fig.1 of refs. \citenum{Kroger_2005} and \citenum{Karayiannis_2009} give a lucid description of this algorithm.
In Fig.~\ref{fig2}b, we plot the resulting $\left\langle Z \right\rangle$ values as a function of $\kappa$. This curve clearly demonstrates the initial increase of $\left\langle Z \right\rangle$ with increasing $\kappa$ and the subsequent drop following $\kappa= 6$, which accompanies the isotropic-nematic transition. Fig.~\ref{fig2}b also shows that $\left\langle Z (\kappa=6)\right\rangle$  exhibits a significant drop from its equilibrium bulk value to its high shear rate value (at $\dot{\gamma} = 0.5$), thereby providing quantitative evidence in favor of our earlier assertion that an increase in shear rate leads to a decrease of interchain entanglements present in the melt and an increased alignment of chains in shear direction.  

We now turn our attention to the rapid increase of $\eta_{\rm GK}$ with increasing $\kappa$ for ring polymers (Fig.~\ref{fig2}a). We attribute this phenomenon to the tendency of semiflexible ring polymers to form cluster phases. We stipulate that with increasing $\kappa$, the degree of cluster formation in our melts increases, which impedes motion under external stimuli and results in the rapid increase of $\eta_{\rm GK}$. We proceed to quantify this clustering behavior as a function of $\kappa$ in Fig.~ \ref{fig2}b. To determine the size of the clusters, we performed a cluster analysis using the density-based spatial clustering of applications with noise (DBSCAN) algorithm \cite{dbscan}. Here, ring polymers are assigned to the same cluster if the distance between their centers-of-mass is smaller than $2.0$. We also took into account the relative orientation between ring polymers by including the weights $\cos(\theta_{ij})^2$, where $\theta_{ij}$ is the angle between the normal vectors $\mathbf{n}_i$ and $\mathbf{n}_j$ of rings $i$ and $j$, respectively ($\mathbf{n}$ is determined as the average cross product between subsequent bond vectors in the ring). We then calculated for each frame the probability distribution $P(S)$ for a polymer to be in a cluster of size $S$, and then defined the mean cluster size as the first moment of $P(S)$, {\it i.e.}, $\langle S \rangle = \sum P(S)S$. Figure~\ref{fig2}b shows the resulting $\langle S \rangle$ as a function of $\kappa$: $\langle S \rangle$ increased monotonically with increasing bending stiffness from $\langle S \rangle \approx 2$ for fully flexible rings up to $\langle S \rangle \approx 17$ for semiflexible rings with $\kappa = 7$. This progression with $\kappa$ is similar to the shear viscosity $\eta$ (cf. Fig.~\ref{fig2}b), suggesting a strong correlation between the clustering of the rings and their rheological properties. Figure~\ref{fig2}b also shows that  $\langle S ( \kappa=7, \dot{\gamma}= 0.5)\rangle$ is significantly smaller than $\langle S ( \kappa=7, \dot{\gamma}= 0)\rangle$, indicating that sheared melts experience a distinct breakdown of the clusters, accompanied by a decrease in the shear viscosity. This phenomenon bears resemblance to the breakdown of entanglements in melts of stiff linear chains subjected to shear, as discussed before. It is worth mentioning at this point that at high shear rates, rings tend to lie in the vorticity-flow (v-f) plane, with their normals pointing towards the gradient direction, so a group of aligned rings may appear stacked to the clustering algorithm (and the resulting $\langle S\rangle$ is overestimated marginally). However, these momentarily ``stacked'' rings slide past each other under shear, so they should not be regarded as clusters of arrested rings which drive up the viscosity.

\begin{figure}[ht!]
    \begin{subfigure}[b]{4in}
      \includegraphics[width=\textwidth]{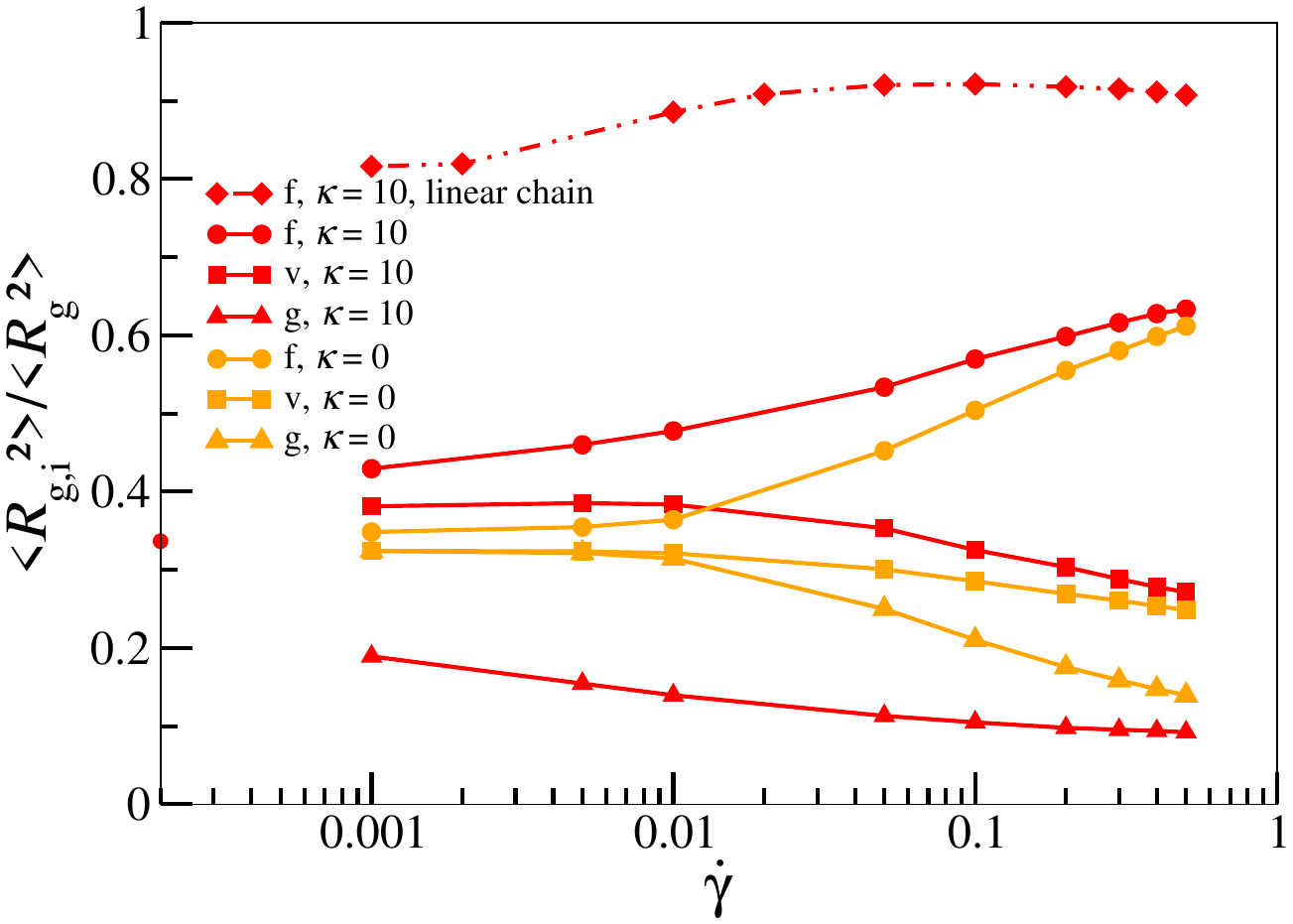}
%      \caption{}
    \end{subfigure}

\caption{$\langle R_{\text{g},i}^2 \rangle / \langle R_\text{g}^2\rangle$ as a function of $\dot{\gamma}$ for $\kappa=0$ and $10$, where $i$ is a placeholder for flow (f), gradient (g) or vorticity (v) directions. (For further explanations see main text.)
Values on the $y$-axis correspond to equilibrium simulations. 
%As there is no preferred orientation in the bulk, the value for the ratio refers to the largest component. 
The curve for $\kappa= 10$ corresponding to linear chains is drawn with dash-dotted lines for reference. All lines serve as guides to the eye only.}
\label{fig3original}
\end{figure}
%\newpage

In Fig.~\ref{fig3original}, we quantify the alignment and deformation of ring polymers in the melt along the direction of shear flow, as a function of shear rate.
For stiffer rings ($\kappa=10$), the ratio of mean squared radii of gyration in flow direction $\langle R_{{\rm g},f}^{2}\rangle / \langle R_{\rm g}^{2}\rangle$ increases from its bulk value (displayed on the $y$-axis) with increasing shear rate, while $\langle R_{{\rm g},g}^{2}\rangle / \langle R_{\rm g}^{2}\rangle$ and $\langle R_{{\rm g},v}^{2}\rangle / \langle R_{\rm g}^{2}\rangle$ decline, the latter to a lesser degree. This indicates an emerging stretching in shear-direction with increasing shear rate and orientation of rings in the f-v plane, which together with the breakup of clusters (for stiffer rings) causes shear-thinning - confirming recent results from Ref.~\citenum{Likos_Liebetreu_2020} obtained for semiflexible rings at lower concentrations. For flexible rings, the trend is qualitatively similar but less pronounced (see Fig.~\ref{fig3original} and Fig. S2 in the Supporting Information, SI).
In comparison, linear chains are aligned much more strongly in the flow direction (see Fig.~\ref{fig3original} and Ref.~\citenum{Datta_Virnau_2021}), as there is no need to orient part of the chain in the vorticity direction. While the conformations of single linear chains under shear are similar to the ones in a sheared melt \cite{Datta_Virnau_2021}, simulations of single rings tend to overestimate the shear-induced deformation as shown in the SI (Fig. S2), which also provides further analysis on stretching as a function of stiffness (Fig. S1).

\section{Segregation of Ring Mixtures under Flow}
\label{sec_3b}

\begin{figure}[ht!]
%\centering
     
      \includegraphics[width=1.3in]{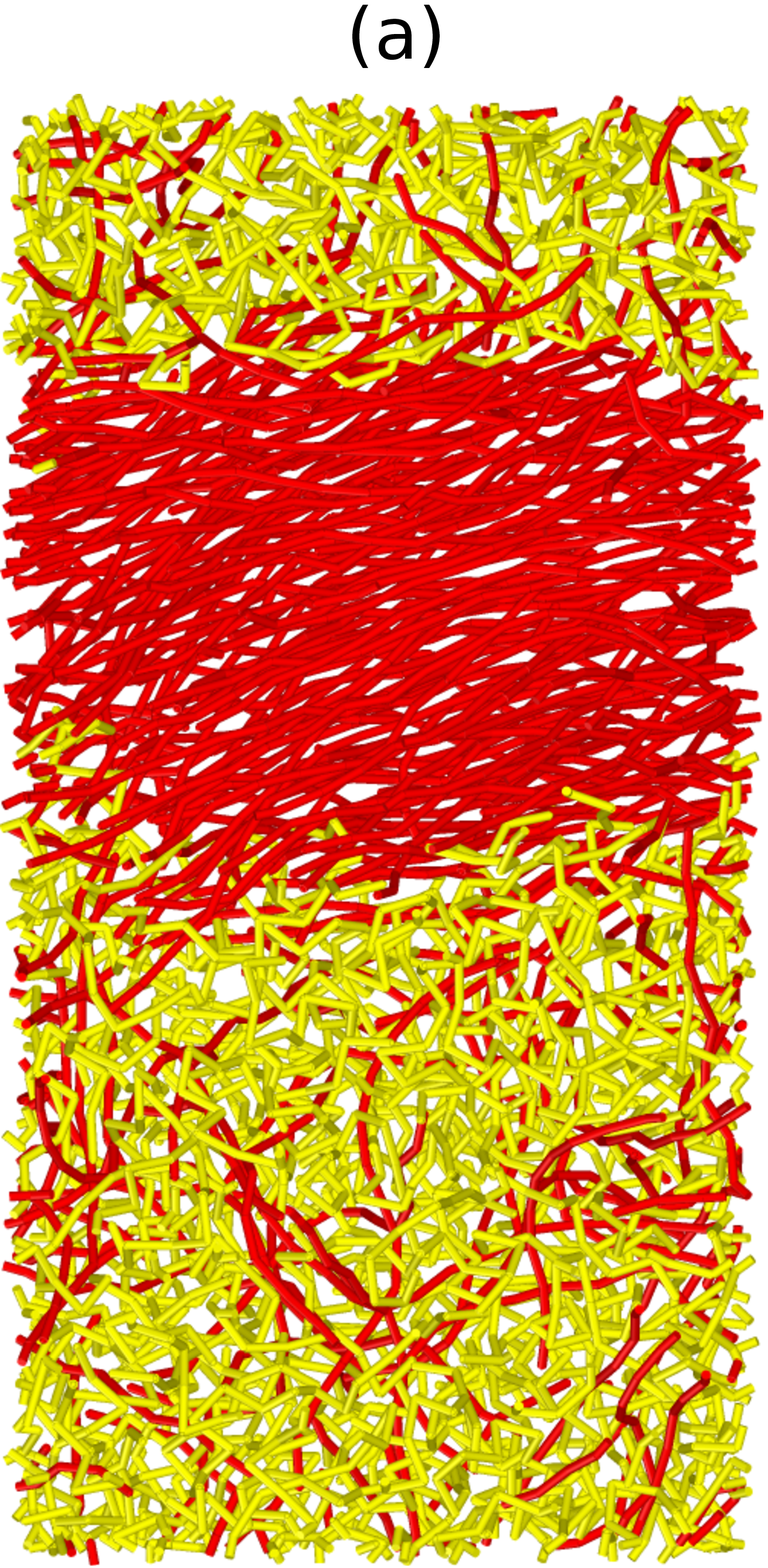}
      \includegraphics[width=1.3in]{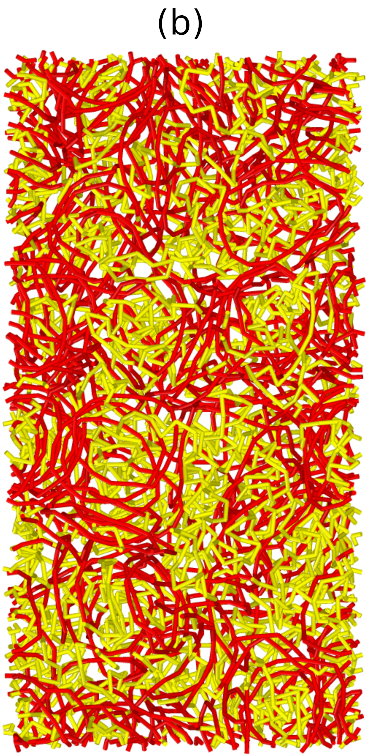}
      \includegraphics[width=3.1in]{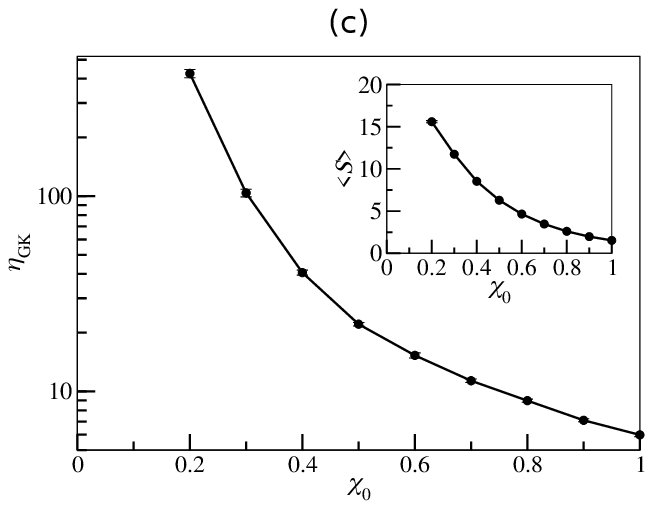}

        \caption{\textbf{(a)} Snapshot of a phase separated binary mixture of linear polymers with stiffness $\kappa=0$ (yellow) and $\kappa=10$ (red) at a fraction of flexible polymers $\chi_0=0.5$. \textbf{(b)} Snapshot of a homogeneous binary mixture of ring polymers with stiffness $\kappa=0$ (yellow) and $\kappa=10$ (red) at $\chi_0=0.5$. \textbf{(c)} Zero-shear viscosity from the Green-Kubo relation $\eta_\text{GK}$ as a function of $\chi_0$ in a binary mixture of ring polymers consisting of flexible and stiff ($\kappa=10$) rings. The inset displays the cluster parameter $\langle S \rangle$ as a function of $\chi_0$. A box size of $15\times15\times15$ was considered. Box dimensions are $15\times30\times15$ for simulations shown in (a) and (b).}
        \label{fig4}
    \end{figure}

In the final section, we turn our attention to binary mixtures of polymers. Figure~\ref{fig4}a shows a 50/50 blend ($\chi_{0}=0.5$) of flexible ($\kappa=0$) and semiflexible ($\kappa=10$) linear polymers at $\rho=0.8$ and $T=1$ ($N=15$ for both the species of polymers). Under these conditions, the blend separates into a nematic phase consisting mostly of stiffer chains ($\kappa=10$) and an isotropic phase mostly made up of flexible chains ($\kappa=0$). In equilibrium, polymers consisting of chemically-identical monomers can separate into two coexisting phases, if their sizes and$/$or stiffness differ significantly \cite{Milchev_2020}. These purely entropic effects cannot be captured by standard Flory-Huggins theory, which disregards the chain connectivity\cite{Kozuch_2016}. Confinement can enhance such segregation effects due to the additional entropic constraints imposed by the confining walls \cite{Nikoubashman_2014_2, Spencer_2022, Matsen_2022}. For the ring mixtures, however, the system remained fully mixed in equilibrium (Fig.~\ref{fig4}b).

In Fig.~\ref{fig4}c, we plot $\eta_\text{GK}$ as a function of the fraction of flexible rings in the blend, $\chi_{0}$. The zero-shear viscosity exhibits a steep decrease with an increasing %fraction of flexible rings 
$\chi_{0}$,which is accompanied by a decrease in $\langle S \rangle$, reiterating the strong correlation between the degree of clustering and viscosity. Note that as long as flexible polymers are in the majority, a substantial increase in clustering accompanied by an exponential increase in viscosity can mostly be impeded.

\begin{figure}[ht!]
%\centering

      \includegraphics[width=1.3in]{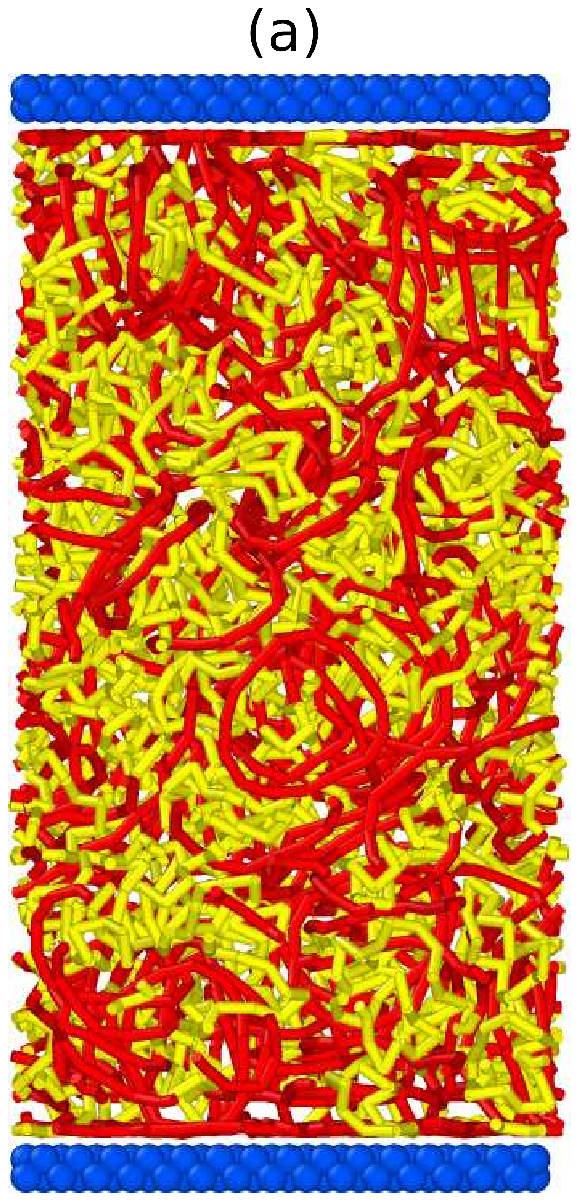}
      \includegraphics[width=1.3in]{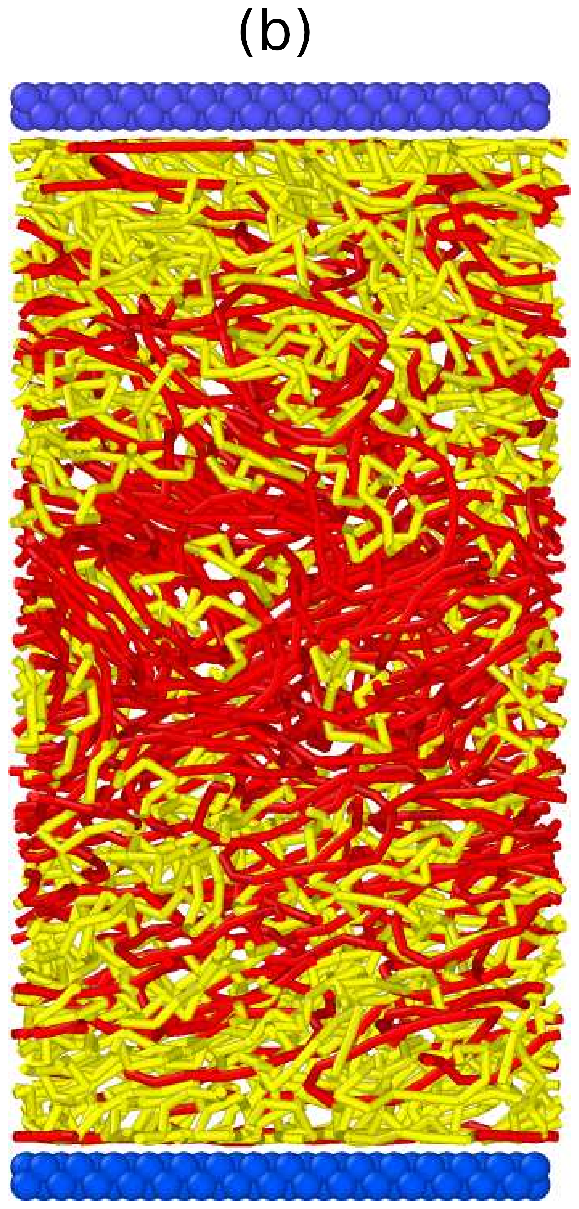}
      \includegraphics[width=3.2in]{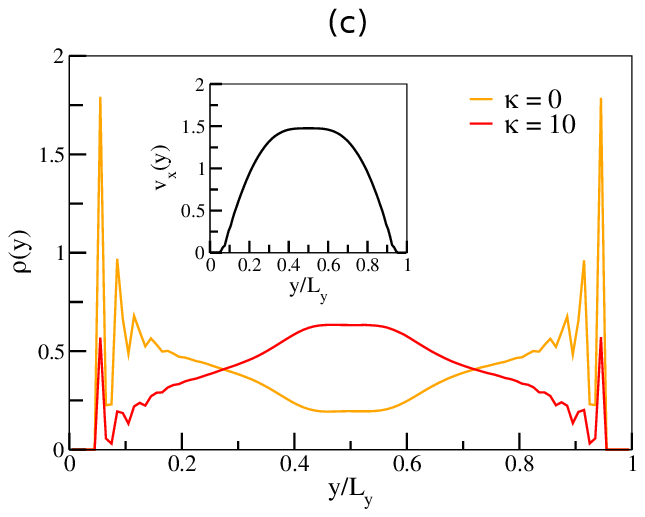}
 
\caption{\textbf{(a)} Equilibrated binary blend of flexible ($\kappa=0$, yellow) and stiffer ($\kappa=10$, red) rings at $\chi_0=0.5$ in a channel confined by particle-based walls (blue). \textbf{(b)} The same blend subject to flow (imposed by applying a constant force $f_x=0.095$ along the $x$-axis to all particles). \textbf{(c)} Density profiles of the respective components along the channel cross-section. The inset shows the profile of the velocity component in flow direction, illustrating that no-slip boundary conditions were attained.}
\label{fig5}
\end{figure}
%\newpage

Next, we investigate channel flow of an evenly composed ($\chi_{0}=0.5$) blend of fully flexible and stiffer ring polymers. Flow is implemented by applying a constant force $f_x=0.095$ on all polymer beads enclosed within confining particle-based walls. The structure of our channel walls and interactions between walls and polymer beads are tuned to impose no-slip boundary conditions. While the blend of rings remains homogeneous in the presence of walls as shown in Fig.~\ref{fig5}a, applying flow leads to a clear segregation into a phase rich in stiff rings located in the low shear central region of the channel, while flexible rings migrate to the high shear region close to the confining walls (Fig.~\ref{fig5}b). Flow-induced segregation of the mixture is further illustrated in Fig.~\ref{fig5}c, where we plot corresponding density profiles of flexible and semiflexible rings alongside the velocity profile. Our simulations thus suggest that stiffness-induced differences in rheological properties may suffice to separate ring polymers of the same mass in microfluidic devices even if they do not phase separate in equilibrium.

The flow-induced separation of flexible and semiflexible rings can be understood by considering the viscoelastic forces, which emerge from the orientation and stretching of the macromolecules along the flow direction during shear \cite{Zhou_2020}. The polymer deformation is smallest in the center of the slit-like channel, where the shear gradient becomes zero, resulting in a cross-stream migration toward the central channel plane. In our setup, \emph{both} polymer components exert elastic forces, which somewhat complicates the theoretical description compared to rigid colloids in viscoelastic media \cite{Tehrani_1996, Nikoubashman_2014, Howard_2015}. Further note that typical theoretical treatments of viscoelastic focusing are based on continuum arguments, which might break down when the constituents of the viscoelastic medium (the flexible rings) and the focused particles (the clusters of stiff rings) become comparable in size. However, previous simulations of star-chain \cite{Srivastava_Nikoubashman_2018} and ring-chain\cite{Weiss_Nikoubashman_2019} mixtures have found a similar focusing of the less deformable species to the center-line, which suggests that this phenomenon occurs even at very small length scales. 

Note that lateral cross channel migration of single polymers in confinement, subject to pressure driven flow has also been studied extensively \cite{Chelakkot_2010, Steinhauser_2012, Utsa_2006, saintillan_2006}. These studies identify hydrodynamic interactions between polymer particles and between polymers and confining walls in addition to shear interactions as the main driving force of cross channel migration. 
In dense systems, however, hydrodynamic interactions are typically suppressed and thus not the primary driving force.
%The tendency of polymers to migrate towards or away from the walls have been thoroughly investigated. Effects of stiffness in polymers on this cross stream migration has also been studied \cite{Chelakkot_2010,Steinhauser_2012} and it has been suggested that \cite{Steinhauser_2012} pressure driven flow can be utilized to separating polymers based on their mechanical properties. However, these studies deal with polymers in dilute solutions, whereas our system of polymer blend is a dense melt where hydrodynamic interactions are suppressed. Hence the segregation that we observe  cannot be explained on the basis of hydrodynamic interactions.

\section{Conclusions}
In this manuscript we have investigated shear thinning in dense melts of flexible and semiflexible ring polymers. We have analyzed how external shear affects microscopic properties like polymer alignment, deformation and cluster formation, thereby changing the macroscopic rheological properties of the melt. Our study focuses particularly on the role of chain stiffness and highlights similarities and differences to systems composed of linear chains \cite{Datta_Virnau_2021}.

For linear polymers, the zero-shear viscosity from the Green-Kubo relation and the low shear rate viscosity show a non-monotonic behavior as a function of chain stiffness with an initial increase, followed by a subsequent drop. While the latter can be attributed to an isotropic-nematic transition, the initial increase at low to moderate stiffness coincides with an increase of interchain entanglements as hypothesized previously in Ref.~\citenum{Datta_Virnau_2021}. 
On the other hand, for melts of ring polymers, the progression of the shear viscosity as a function of stiffness is quite different. With increasing $\kappa$ and ring size, there is a steep and monotonic increase in viscosity, which can be attributed to the rapid increase in cluster size in the system \cite{Bernabei_2013, Slimani_2014, Poier_2015, Poier_2016, Likos_Liebetreu_2020}. To quantify this effect, we have proposed a novel measure, $\langle S \rangle$, to determine the degree of clustering. With increasing shear rate, clusters dissolve, rings are stretched in the flow direction and oriented in the vorticity-flow plane, resulting in pronounced shear thinning \cite{Likos_Liebetreu_2020}.

In the final section of our work, we have simulated blends of flexible ($\kappa=0$) and stiffer ($\kappa=10$) rings. Unlike blends of stiff and flexible linear chains, ring blends do not phase separate in equilibrium, but form a homogeneous phase even in the presence of walls. We establish that semiflexible and flexible rings can be separated in a channel flow scenario due to differences in their intrinsic rheological properties. Flow-induced segregation \cite{Weiss_Nikoubashman_2019, Srivastava_Nikoubashman_2018} could thus be used in microfluidics to separate rings of similar mass, size and chemical composition according to stiffness.

\section*{Acknowledgements}
This work was funded by the Deutsche Forschungsgemeinschaft (DFG, German Research Foundation - SFB TRR 146, Project number 233530050 (projects B2, C1, and C5). A.N. acknowledged further funding from the DFG through Project No. 470113688. The authors are also immensely thankful for the computation time granted on the HPC cluster Mogon at Johannes Gutenberg University, Mainz. The authors also acknowledge the valuable inputs provided by Dr. Leonid Yelash, Prof. Maria Luk\'a\v{c}ov\'a-Medvidov\'a and Mr. Andreas Schömer.

%\section{Summary and Outlook}
%\label{sec_6}

%\section*{Acknowledgements}

%%%%%%%%%%%%%%%%%%%%%%%%%%%%%%%%%%%%%%%%%%
%% Optional
%\appendixtitles{yes} % Leave argument "no" if all appendix headings stay EMPTY (then no dot is printed after "Appendix A"). If the appendix sections contain a heading then change the argument to "yes".
%\appendixstart
%\appendix

%\newpage
%%%%%%%%%%%%%%%%%%%%%%%%%%%%%%%%%%%%%%%%%%
%\end{paracol}
%%%%%%%%%%%%%%%%%%%%%%%%%%%%%%%%%%%%%%%%%%

\newpage
%\reftitle{References}
%\externalbibliography{yes}
\bibliography{cit1.bib}

\end{document}

\newcommand{\beginsupplement}{%
        \setcounter{table}{0}
        \renewcommand{\thetable}{S\arabic{table}}%
        \setcounter{figure}{0}
        \renewcommand{\thefigure}{S\arabic{figure}}%
     }

\beginsupplement

\section{Stretching of rings}
\begin{figure}[ht!]
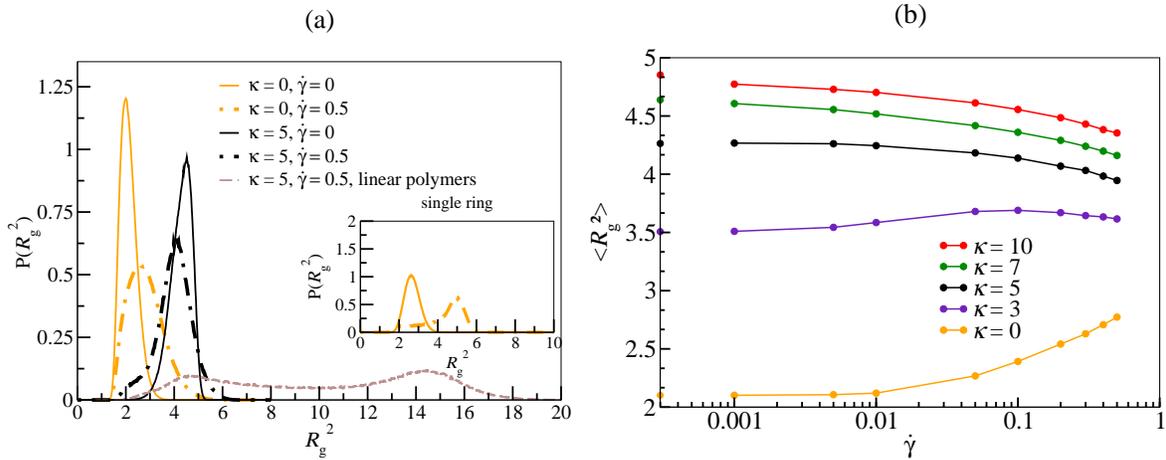

      \includegraphics[width=3in]{S1A_Final.eps}
      \includegraphics[width=3.05in]{S1B_Final.eps}

\caption{\textbf{(a)} Probability distributions P($R_\text{g}^{2}$) for flexible chains ($\kappa$ = 0) and semiflexible chains ($\kappa=5$) having shear rates $\dot{\gamma}=0$ and $0.5$, with P($R_\text{g}^{2}$) for $\kappa= 5$ corresponding to linear chains drawn with dotted lines for reference. The inset shows P($R_\text{g}^{2}$) for flexible single rings in equilibrium and at shear rate $\dot{\gamma}=0.5$.
\textbf{(b)} Mean-squared radius of gyration  $\langle R_\text{g}^2\rangle$ as a function of $\dot{\gamma}$ for $\kappa=0, 3, 5,7$ and $10$. Values on the $y$-axis (in \textbf{a} and \textbf{b}, color scheme like in Fig. 1b of the main text) correspond to equilibrium simulations without shear. All lines serve as guides to the eye. }
\label{fig6}
\end{figure}

In Fig.~\ref{fig6}, we study the stretching of individual chains present in the melt under shear. In Fig.\ref{fig6}a, we compare the distribution of the square of the radius of gyration, $R_\text{g}^{2}$ between rings of different stiffness under equilibrium and shear. While the distribution of $R_\text{g}^{2}$ pertaining to  flexible rings is narrow and centered around a small value of $R_\text{g}^{2}$, i.e., rings are rather compact, under a high shear rate of $\dot{\gamma} = 0.5$, the distribution develops an extended tail to the right which indicates that chains undergo shear induced stretching. This also explains the rise in $R_\text{g}^{2}$ as function of $\dot{\gamma}$ for flexible chains (Fig.~\ref{fig6}b). However, stiffer rings are already stretched in equilibrium and the effect of the applied shear is numerically very small. It is interesting to note that for the case of semiflexible ($\kappa=5$) linear chains, at high shear rate the distribution of $R_\text{g}^{2}$ shows two distinct peaks corresponding to two distinct movement modes, namely stretching and tumbling \cite{nikoubashman:mm:2017, Datta_Virnau_2021, Kong2019}. High shear leads to the abundance of tumbling U shaped conformations in addition to the stretched conformations for semiflexible chains, leading to the emergence of the second peak in the $R_\text{g}^{2}$ distribution. However, for semiflexible rings having  $\kappa=5$ high shear does not lead to the emergence of any such new peak in the $R_\text{g}^{2}$ distribution. Only the distribution at $\dot{\gamma} = 0.5$  develops a tail to the left (Fig.~\ref{fig6}a.)

\section{Comparison to single ring simulations}

In order to assess the role of cooperativity effects, we have simulated single isolated ring polymers in an external shear profile. The bonded and nonbonded potentials were the same as in the melt simulations. The equation of motion for the monomer $i$ is

\begin{equation}
    \dot{\mathbf{r}_i} = \frac{1}{\zeta} \mathbf{f}_i + \dot{\gamma} z_i \mathbf{e}_x + \mathbf{\xi}_i(t),
\end{equation}

with $\zeta$ being the monomeric friction, $\mathbf{f}_i$ the conservative force derived from the intermolecular potentials, $\mathbf{e}_x$ a unit vector pointing in the flow direction, and $\xi_{i,\alpha}(t)$ a stochastic force. The gradient direction (g) is $z$ and the vorticity direction (v) is $y$. The stochastic force is given by an uncorrelated Gaussian 
white noise with $\langle \xi_{i \alpha}(t) \rangle \equiv 0$ and
$\langle \xi_{i \alpha}(t) \xi_{i \beta}(t') \rangle = (2 k_\text{B} T /\zeta) \:  \delta_{ij} \delta_{\alpha \beta} \delta(t-t')$. The equations of motion were integrated using an Euler forward algorithm with  time step $\Delta t = 10^{-4} \zeta \sigma^2/\epsilon$. The time scales of the two models were adjusted by the same procedure as described in Ref. \citenum{Datta_Virnau_2021}.

\begin{figure}[ht!]
    \begin{subfigure}[b]{3.5in}
      \includegraphics[width=\textwidth]{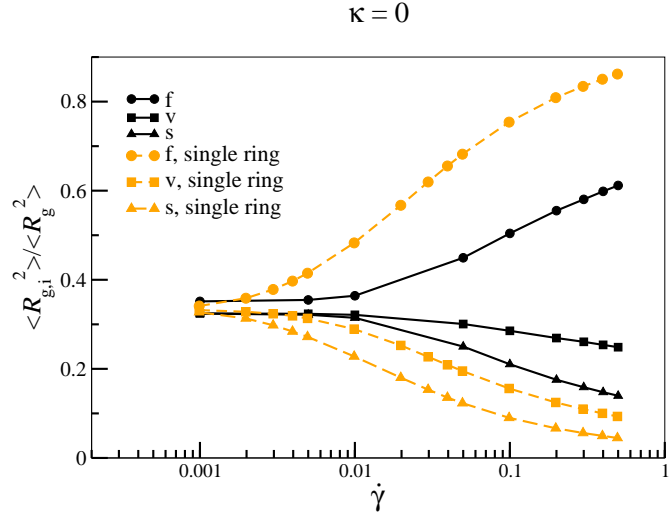}
%      \caption{}
    \end{subfigure}

\caption{ Ratios of the flow (f), gradient (g) and vorticity (v) components of the mean-squared radius of gyration, $\langle R_\text{g,f}^2 \rangle$, $\langle R_\text{g,g}^2 \rangle$, and $\langle R_\text{g,v}^2 \rangle$ relative to $\langle R_\text{g}^2\rangle$ as a function of $\dot{\gamma}$ for melt (solid lines) and single ring (dashed lines) simulations at $\kappa=0$.}
\label{fig7}
\end{figure}

 The qualitative behavior of isolated rings is similar to that observed for the rings in the melt, i.e., the rings are stretched in the flow direction and compressed in both gradient and vorticity direction. However, quantitatively, isolated chains are much more strongly affected by the shear flow than melt chains.
 The inset of  Fig.~\ref{fig6}b clearly shows that for flexible rings subject to a high shear rate, the resulting  $\langle R_\text{g,g}^2\rangle$ distribution for single rings exhibits a peak at higher value of $R_\text{g,g}^2$ as compared to that of the melt of flexible rings. This behavior indicates that single rings exhibit a larger degree of chain stretching than chains in melt, when both are subject to high shear. Figure~\ref{fig7} also reveals that at higher shear rates the magnitude of alignment in the direction of shear is considerably larger for single rings. So responses of rings pertaining to single ring simulations to shear are qualitatively similar but quantitatively very different from those of rings pertaining to melt simulations.

\newpage

\bibliography{cit1.bib}